\def\eqi{\begin{equation}}
\def\eqf{\end{equation}}
\def\eqia{\begin{eqnarray}}
\def\eqfa{\end{eqnarray}}
\def\rp#1#2{{#1\over#2}}

\def\rfr#1{eq.(\ref{#1})}
\def\rfrs#1#2{eqs.(\ref{#1})-(\ref{#2})}

\def\Rfr#1{Eq.(\ref{#1})}

\def\rp#1#2{{#1\over#2}}

\def\lb#1{\label{#1}}

\def\bm#1{{\mbox{\boldmath$#1$\unboldmath}}}

\def\X{\Xi}

\def\to{$T_{\rm obs}$}

\documentclass[onecolumn]{aa} % For Astronomy and Astrophysics

\begin{document}

\title{New ways in testing  post-Newtonian gravity in the Solar System scenario from planetary motion analysis}

\author{L. Iorio\inst{1}}

\offprints{L. Iorio}

\institute{Dipartimento di Fisica dell'Universit${\rm \grave{a}}$
di Bari, Via Amendola 173, 70126, Bari, Italy\\
\email{Lorenzo.Iorio@ba.infn.it} }

\date{Received , 2004; accepted , 2004}

\abstract{In this chapter we first calculate the post--Newtonian
gravito-electric secular rate of the mean anomaly of a test
particle freely orbiting a spherically symmetric central mass.
Then, we propose a novel approach to suitably combine the
presently available planetary ranging data to Mercury, Venus and
Mars in order to determine, simultaneously and independently of
each other, the Sun's quadrupole mass moment $J_{2\odot}$ and the
post-Newtonian gravito-electric secular advances of the perihelion
and the mean anomaly. This would also allow to obtain the PPN
parameters $\gamma$ and $\beta$ independently. We propose to
analyze the time series of three linear combinations of the
observational residuals of the rates of the nodes $\dot\Omega$,
the longitudes of perihelia $\dot\pi$ and mean anomalies
$\dot\mathcal{M}$ of Mercury, Venus and Mars built up in order to
account for the secular precessions induced by the solar
oblateness and the post-Newtonian gravito-electric forces.
According to the present-day EPM2000 and DE405 ephemerides
accuracy, the obtainable precision would be of the order of
10$^{-4}$--10$^{-5}$ for the PPN parameters and, more
interestingly, of $10^{-9}$ for $J_{2\odot}$. Second, we
investigate the possibility of measuring the post-Newtonian
Lense-Thirring secular effect induced by the gravito-magnetic
field of the Sun on the planetary orbital motion by analyzing two
linear combinations of the residuals of the nodes $\Omega$ of
Mercury, Venus and Mars. The proposed combinations are, by
construction, unaffected by the quadrupole mass moment $J_2$ of
the Sun's gravitational potential. Moreover, they are insensitive
also to the post-Newtonian gravito-electric field of the Sun
because it affects only the perihelia $\omega$ and the mean
anomalies $\mathcal{M}$. The obtainable observational accuracy in
the proposed measurement would be of the order of 19$\%$ and
36$\%$. However, it must be pointed out that a major source of
systematic error could be represented by the residual mismodelled
classical $N-$body secular precessions. The second proposed
combination is insensitive just to such aliasing effect.

}

\titlerunning {The solar gravito-magnetic field and its possible measurement}

\maketitle

\section{Introduction}
In this chapter we wish to investigate the possibility of
performing some tests of Newtonian and post-Newtonian gravity in
the Solar System arena extending to it some approaches which, in
recent years, have proven to be fruitful in certain
Earth-artificial satellites systems. More precisely, we will deal
with certain classical and post-Newtonian gravito-electromagnetic
secular effects affecting the longitudes of the ascending nodes
$\Omega$, the longitudes of the perihelia $\pi$ and the mean
anomalies $\mathcal{M}$ of the Solar System's planets. The
proposed tests could be done in a relatively simple and cheap way
(with respect to new missions involving the construction and
launch of one or more spacecrafts); moreover, they would also
benefit of the improvements in the planetary ephemerides coming
from future planned space missions like Messenger and BepiColombo.
The obtainable accuracies are comparable with those of future
proposed or planned expensive and complex interplanetary missions.
In the case of the gravito-magnetic field, it would be the first
attempt to measure its effect on the planetary orbital motion in
the gravitational field of the Sun.

Historically, one of the first tests of the Einstein General
Theory of Relativity (GTR) was the successful explanation of the
anomalous secular gravito-electric perihelion advance of Mercury
in the gravitational field of Sun (Einstein 1915). On the
contrary, another prediction of the Einstein's theory of
gravitation for the motion of test particles, i.e. the much
smaller gravito-magnetic Lense-Thirring effect (Lense \& Thirring
1918), is still awaiting for an undisputable and direct
observational check. As we will see, these post-Newtonian features
of the planetary motion are strictly connected with the problem of
the quadrupole mass moment $J_{2\odot}$ of Sun. In (Pireaux and
Rozelot 2003) a theoretical range $J_{2\odot}=(2\pm 0.4)\times
10^{-7}$ is admitted for the Sun's oblateness. For this topic and
the interplay between this effect and the gravito-electric
Einstein precession see the recent review (Pireaux \& Rozelot
2003) and the references therein and (Ciufolini \& Wheeler 1995).
Basically, the point is the following. In regard to the secular
orbital motions, to which we are interested here, the Sun's mass
quadrupole moment induces classical effects which have,
qualitatively, the same temporal signature of the relativistic
ones. This means that they could corrupt the recovery of the
genuine post-Newtonian features of motion because they could not
be removed from the time series without removing the
post-Newtonian signal of interest as well. All depends on the
precision with which $J_{2\odot}$ is known: should the
mismodelling in it induce classical residual precessions larger
than the relativistic ones, there would  be no hope to get a
reliable test of relativistic gravity. As we will see later, the
same problems could come from the $N-$body secular precessions. In
this chapter we propose to disentangle such effects by measuring
them in an independent way. More precisely, we propose to extend a
certain approach used in Earth artificial satellite motion
analysis to the interplanetary arena in order to single out just
certain post--Newtonian (and Newtonian) orbital motion features
independently of the Parameterized--Post--Newtonian (PPN)
framework (Will 1993) which is usually employed in testing
competing metric theories of gravity. Instead, the current
approach consists of testing the post--Newtonian equations of
motion as a whole in terms of the PPN parameters which are
determined from multi--parameters fits together with other
astrodynamical quantities.

In Table \ref{param} the orbital parameters of the inner planets
of the Solar System are reported.
\begin{table}
  \caption[]{Orbital parameters of Mercury, Venus and Mars
(\texttt{http://nssdc.gsfc.nasa.gov/planetary/factsheet/}). For
the Astronomical Unit (A.U.) we use the value 1 A.U.=149597870691
m of the DE405 (Standish 1998) and EPM2000 ephemerides (Pitjeva
2001a; 2001b). The angle $\epsilon$ refers to the inclination to
the ecliptic.}
     \label{param}
$$
\begin{array}{p{0.5\linewidth}lll}
            \hline
            \noalign{\smallskip}
Planet  & a\ (A.U.) & \epsilon\ (^\circ) & e \\
\noalign{\smallskip}
            \hline
            \noalign{\smallskip}
Mercury & 0.38709893 &  7.00487 & 0.20563069 \\
Venus & 0.72333199 & 3.39471 & 0.00677323 \\
Mars & 1.52366231 & 1.85061 & 0.09341233 \\
\noalign{\smallskip}
            \hline
         \end{array}
     $$
\end{table}

\section{The post-Newtonian effects}
\subsection{The gravito-electric effects}
In the framework of the standard PPN formalism the post-Newtonian
gravito-electric acceleration induced by the Schwarzschild--like
part of the spacetime metric and experienced by a test body freely
falling around a static, spherically symmetric central mass $M$ is
(Soffel 1989) \eqi \bm a_{\rm GE}=\rp{GM}{c^2
r^3}\left\{\left[2(\gamma+\beta)\rp{GM}{r}-\gamma(\bm v\cdot\bm v)
\right]\bm r+2(1+\gamma)(\bm r\cdot\bm v)\bm
v\right\},\lb{age}\eqf where $G$ is the Newtonian gravitational
constant, $c$ is the speed of light in vacuum, $\bm r$ and $\bm v$
are the position and velocity vectors, respectively, of the test
body, $\gamma$ and $\beta$ are the standard
Eddington-Robertson-Schiff PPN parameters ($\gamma=\beta=1$ in
GTR; in this case \rfr{age} reduces to the expression of (Mashhoon
et al. 2001)). Note that $r$ here is the standard isotropic radial
coordinate, not to be confused with the Schwarzschild radial
coordinate $r^{'}=r[1+GM/(2c^2 r)]^2$. In the usual orbital data
reductions it is just $r$ which is employed.

By considering it as a small perturbation to the Newtonian
monopole acceleration it is possible to work out its effect on the
orbital motion of  a test body with the standard perturbative
techniques. The secular rate of the argument of pericentre is
given by the well known formula \eqi\dot\omega_{\rm
GE}=\frac{3nGM}{c^2 a(1-e^2)}\nu_{\rm GE },\lb{ge}\eqf where
 \eqi\nu_{\rm GE }=\frac{2+2\gamma-\beta}{3},\lb{nuge}\eqf
$a$ and $e$ are the semimajor axis and the eccentricity,
respectively, of the test particle's orbit and $n=\sqrt{GM/a^3}$
is the (unperturbed) Keplerian mean motion. The orbital period of
an unperturbed, two-body Keplerian ellipse is $P=2\pi/n$. Note
that \rfr{ge} is an exact result valid to all order in $e$.

As we will show, also the mean anomaly $\mathcal{M}$ is affected
by a post-Newtonian gravito-electric secular rate. The Gauss
perturbative equation for $\mathcal{M}$ is given by
\eqi\rp{d\mathcal{M}}{dt}=n-\rp{2}{na}A_R\left(\rp{r}{a}\right)-\sqrt{1-e^2}\left(\rp{d\omega}{dt}+\cos
i\rp{d\Omega}{dt} \right),\lb{manom}\eqf where $\Omega$ and $i$
are the longitude of the ascending node and the inclination of the
test particle's orbit to the equator of the central mass,
respectively, and $A_R$ is the radial component of the perturbing
acceleration. In order to obtain the post-Newtonian
gravito-electric secular rate of the mean anomaly one must
consider the radial component of \rfr{age}, insert it in the
second term of the right-hand-side of \rfr{manom}, evaluate it on
the unperturbed Keplerian ellipse, characterized by
$r=a(1-e^2)/(1+e\cos f)$ and by the radial and along-track
components of the velocity vector which are $v_{R}=nae\sin
f/\sqrt{1-e^2}$, $v_{T}=na(1+e\cos f)/\sqrt{1-e^2}$, respectively
($f$  is the orbiter's true anomaly),  multiplying it by
\eqi\rp{dt}{P}=\rp{(1-e^2)^{3/2}df}{2\pi(1+e\cos f )^2},\eqf  and
integrating over one orbital revolution, i.e. from 0 to $2\pi$. It
turns out that the post-Newtonian gravito-electric secular rate of
the mean anomaly is, to order $\mathcal{O}(e^2)$
\eqi\dot\mathcal{M}_{\rm GE}\sim-\rp{nGM}{c^2
a}\left[\left(2+4\gamma+3\beta\right)\left(1+\rp{e^2}{2}\right)+\left(2+\gamma\right)e^2\right].\lb{manomi}\eqf
We can define \eqi\mu_{\rm GE}=\rp{2+4\gamma+3\beta}{9}.\eqf It
turns out that the second term of \rfr{manomi} induces for Mercury
an additional shift of 1.789 arcseconds/century ($''$ cy$^{-1}$ in
the following), while for the other planets it is of the order of
$10^{-2}$--$10^{-4}$ $''$ cy$^{-1}$. Thus, for them the secular
rate of the mean anomaly can be written as
\eqi\dot\mathcal{M}_{\rm GE}\sim-\rp{9nGM}{c^2
a\sqrt{1-e^2}}\mu_{\rm GE}.\eqf However, as we will see later, the
mean anomaly of Mercury will not be used in the combined residuals
strategy outlined in the following.
For Mercury, the post-Newtonian gravito-electric effect induced by
the Sun on $\omega$ and $\mathcal{M}$, according to GTR, amounts
to 42.980 $''$ cy$^{-1}$ and -127.949 $''$ cy$^{-1}$,
respectively. The most accurate estimate of the gravito-electric
perihelion advance seems to be that obtained for Mercury by E.M.
Standish in 2000 with the DE405 ephemerides (Standish 1998) and
reported in (Pireaux and Rozelot 2003). He averaged the Mercury's
perihelion evolution over two centuries by using the DE405
ephemerides with and without the post-Newtonian accelerations.
Standish included in the force models also the solar oblateness
with $J_{2\odot}=2\times 10^{-7}$, so that the residuals for
Mercury accounted for the post-Newtonian effects only; the
determined shift was $42.98\pm 0.0023$ $''$ cy$^{-1}$. The same
approach for the mean anomaly (Standish E M private communication)
has yielded to $-130.003\pm 0.0027$ $''$ cy$^{-1}$ (See also Table
\ref{rates} later). Note that the quoted uncertainty do not come
from direct observational errors. They depend on the fact that in
the force models used in the numerical propagation many
astrodynamical parameters occur (masses of planets, asteroids,
etc.); their numerical values come from multiparameter fits of
real data and, consequently, are affected by observational errors.
Such numerical tests say nothing about if GTR is correct or not;
they just give an idea of what would be the obtainable accuracy
set up by our knowledge of the Solar System arena if the Einstein
theory of gravitation would be true.

In regard to the possibility of constructing time series of
planetary mean anomalies, it must be noted that a certain effort
would be required. Indeed, so far this orbital elements has never
been utilized, so that the partials, e.g., should be computed
(Pitjeva, private communication 2004).
\subsection{The gravito-magnetic Lense-Thirring effect}
Another post-Newtonian secular precession which affects not only
$\omega$ but also $\Omega$ is the Lense-Thirring effect (Lense \&
Thirring 1918) induced by the proper angular momentum $J$ of the
central body \eqia\dot\omega_{\rm LT}&=&\frac{-6GJ\cos i}{c^2 a^3
(1-e^2)^{3/2}}\mu_{\rm LT},\lb{gm}\\
\dot\Omega_{\rm LT}&=&\frac{2GJ}{c^2 a^3 (1-e^2)^{3/2}}\mu_{\rm
LT},\lb{gm2}\eqfa where \eqi\mu_{\rm LT}=\frac{1+\gamma}{2}.\eqf
By assuming for the Sun $J_{\odot}=1.9\times 10^{48}$ g cm$^2$
s$^{-1}$ (Pijpers 2003), the gravito-magnetic effect on, e.g.,
Mercury's perihelion is of the order of $10^{-3}$ $''$ cy$^{-1}$
(De Sitter 1916). Such small value is at the edge of the present
sensitivity (0.002 $''$ cy$^{-1}$) in determining the Mercury
perihelion shift from the ephemerides (cfr. the results obtained
by Standish). Moreover, it should be considered that, even if
future improvements of the obtainable observational sensitivity in
interplanetary ranging allowed to consider the possibility of
measuring the Lense-Thirring perihelion advance of Mercury, the
impact of the systematic errors due to the uncertainties in the
solar oblateness would severely limit the realistic accuracy
obtainable in such demanding measurement. Indeed, by assuming an
uncertainty of $\sigma_{J_{2\odot}}=0.4\times 10^{-7}$, the error
in the secular precession induced by the Sun's quadrupole moment
would amount to 245$\%$ of the Lense-Thrring shift of $\pi^{\rm
Merc}$ (see below for the definition of $\pi$). Another important
source of systematic error in the measurement of such a tiny
effect would be represented by the $N-$body classical secular
precessions which are of the order of $10^2-10^3$ $''$ cy$^{-1}$
(see on the WEB
\texttt{http://ssd.jpl.nasa.gov/elem$\_$planets.html}). Indeed,
their residual mismodelled part could severely bias the recovery
of the Lense-Thirring effect. In this chapter we will propose a
possible approach to overcome these problems. It must be noted
that, if, on the other hand, we look at a Lense--Thirring test as
a way to measure the Sun's angular momentum by assuming the
validity of GTR, a measurement of the solar gravito-magnetic field
would have a significance only if the obtainable accuracy was
better than 10$\%$; indeed, among other things, the present-day
uncertainty in the Sun's angular momentum $J_{\odot}$, which could
be measured from the Lense-Thirring precessions, is just of the
order of 10$\%$ (Xu \& Ni 1997) in various solar models
(Patern\'{o} et al. 1996; Elsworth et al. 1995) or even less in
the framework of asteroseismogyrometry (Pijpers 2003).

At present, the only performed attempts to explicitly extract the
Lense-Thirring signature from the data of orbiting masses in the
Solar System are due to Ciufolini and coworkers who analyzed the
laser-ranged data of the orbits of the existing LAGEOS and LAGEOS
II Earth artificial satellites (Ciufolini et al. 1998). A
20$-30\%$ precision level in measuring the terrestrial
gravito-magnetic field is claimed, but other scientists judge
these evaluations too optimistic and propose different error
budget (Ries et al. 2003). In April 2004 the GP-B spacecraft has
been launched. It will carry out a very complex and challenging
mission which should be able to measure another gravito-magnetic
effect, i.e. the precession of the spins of four superconducting
gyroscopes (Schiff 1960) carried on board at a claimed accuracy of
1\% or better (Everitt et al. 2001). Almost twenty years ago it
was proposed to launch a third LAGEOS-like satellite$-$the LAGEOS
III/LARES$-$ and to analyze the time series of the sum of the
residuals of the nodes of LAGEOS and LARES (Ciufolini 1986) or
some other combinations of residuals of the nodes and the perigees
of LARES and both the existing LAGEOS satellites (Iorio et al.
2002a). The obtainable accuracy would probably be of the order of
1$\%$ too. Mainly funding problems have prevented, up to now, from
implementing such relatively easy and cheap mission. Recently, the
possibility of measuring the Lense-Thirring precessions  by means
of the Relativity-dedicated OPTIS spacecraft, which could be
launched in the same orbital configuration of LARES, has been
considered (Iorio et al., 2004). The recently proposed LATOR
(Turyshev et al. 2004) and ASTROD (Ni et al. 2004) missions would
be sensitive to the gravito-magnetic part of the bending of light
rays and time delay in the gravitational field of the Sun.
Finally, it must be noted that, according to K. Nordvedt Jr.
(Nordvedt 2003), the multidecadal analysis of the Moon's motion
with the Lunar Laser Ranging (LLR) technique strongly supports the
existence of the gravito-magnetic force\footnote{According to
Nordvedt (2003), the Earth--Moon range is affected by
long--periodic harmonic perturbations of gravito-magnetic origin
whose amplitudes are of the order of 5 m and the periods are
monthly and semi-monthly. The amplitudes of the lunar motion at
both these periods are determined to better than half a centimeter
precision in the total orbital fit to the LLR data. } as predicted
by GTR, although in an indirect way. ``It often has been claimed
that the presence of gravitomagnetism within the total
gravitational interaction has not been experimentally confirmed
and measured. Indeed, different experiments have been under
development to explicitly observe the effects of this historically
interesting prediction of general relativity. But this
gravito-magnetic acceleration already plays a large role in
producing the final shape of the lunar orbit, albeit in
conjunction with the rest of the total equation of motion; the
precision fit of the LLR data indicates that gravitomagnetism's
presence and specific strength in the equation of motion can
hardly be in doubt. [...] It would be impossible to understand
this fit of the LLR data without the participation of the
gravito-magnetic interaction in the underlying model, and with
strength very close to that provided by general relativity,
$\gamma=1$".
\section{The solar oblateness}
The solar quadrupole mass moment $J_{2\odot}$ is an important
astrophysical parameter whose precise knowledge could yield many
information about the inner structure and dynamics of our star. A
reliable evaluation of $J_{2\odot}$ still faces some controversy:
on one side, the theoretical values strongly depend on the solar
model used, whereas accurate measurements are very difficult to
obtain from observations. For all this matter see the recent
review (Pireaux \& Rozelot 2003) and (Rozelot et al. 2004). From
an observational point of view, $J_{2\odot}$ is not directly
accessible. In this context a dynamical determination of
$J_{2\odot}$, analyzing, e.g., the orbits of the inner planets of
the Solar System, is interesting because it might be compared with
those derived from solar model dependent values of the oblateness.
However, it is not simple to reach this goal because of the
interplay between the effects of the solar quadrupole moment with
those induced by the post-Newtonian gravito-electromagnetic
forces. Instead of the trajectory of planets, it would be possible
to infer $J_{2\odot}$ from accurate tracking of some drag-free
spacecraft orbiting within a few radii of the solar center. This
will be the approach followed by, e.g. the BepiColombo mission
(see Section \ref{newmiss}). Alternatively, the Sun's quadrupole
mass moment can be inferred from in-orbit measurement of solar
properties, like the SOHO-$MDI$ space-based observations
(Armstrong \& Kuhn 1999), or from Earth-based observations like
those realized, e.g., with the scanning heliometer of the Pic du
Midi Observatory (Rozelot et al. 2004).
\subsection{The classical precessions induced by the solar oblateness}
For a given planet of the Solar System orbiting the Sun, apart
from the classical effects induced by the precession of the
equinoxes and by the other planets and major asteroids which are
routinely accounted for in the ephemerides computations (Pitjeva
2001a; 2001b), the oblateness of Sun induces also secular
precessions on $\Omega,\ \pi=\omega+\Omega\cos i$ and
$\mathcal{M}$ given by \eqia
\dot\Omega_{J_{2\odot}}&=&-\frac{3}{2}\frac{nJ_{2\odot}}{(1-e^2)^2}\left(\frac{R_{\odot}}{a}\right)^2\cos i,\\
\dot\pi_{J_{2\odot}}&=&-\frac{3}{2}\frac{nJ_{2\odot}}{(1-e^2)^2}\left(\frac{R_{\odot}}{a}\right)^2\left(\frac{3}{2}\sin^2
i-1 \right),\lb{obla}\\\lb{oblaanom}
\dot\mathcal{M}_{J_{2\odot}}&=&\frac{3}{4}\frac{nJ_{2\odot}
}{(1-e^2)^{3/2}}\left(\frac{R_{\odot}}{a}\right)^2 (3\cos^2 i-1),
\eqfa The explicit expressions for the secular precessions of
$\Omega$ and $\omega$ up to degree $\ell=20$ can be found in
(Iorio 2003). For Mercury\footnote{As pointed out in (Milani et
al., 2002), the angle $i$ refers to the inclination between the
planet's orbital plane and the fixed reference plane of the
celestial reference frame; it is not the angle $\epsilon$ between
the planet's orbital plane and the ecliptic. It turns out that
$i\sim\epsilon/2$. For Mercury $\epsilon=7.00487^\circ$. } the
Newtonian precessions due to Sun oblateness, with
$J_{2\odot}=2\times 10^{-7}$, are of the order of $10^{-2}$ $''$
cy$^{-1}$ for $\Omega$, $\pi$ and $\mathcal{M}$.
\section{The interplay between the solar oblateness and the
post-Newtonian precessions}
In Table \ref{tab1} the relevant parameters of the classical and
post-Newtonian gravito-electric secular precessions of the nodes,
the perihelia and the mean anomalies of Mercury, Venus and Mars
are reported.
\begin{table}
\caption[]{Post-Newtonian gravito-electric precessions and
coefficients of Newtonian precessions of the node, the perihelion
and the mean anomaly for Mercury, Venus and Mars in $''$
cy$^{-1}$. The values $\dot\pi_{\rm GE}$, and
$\dot\mathcal{M}_{\rm GE}$ are calculated with GTR. The
coefficients $\dot\Omega_{.2}$, $\dot\pi_{.2}$ and
$\dot\mathcal{M}_{.2}$ are
$\partial(\dot\Omega_{J_{2\odot}})/\partial(J_{2\odot})$,
$\partial(\dot\pi_{J_{2\odot}})/\partial(J_{2\odot})$ and
$\partial(\dot\mathcal{M}_{J_{2\odot}})/\partial(J_{2\odot})$,
respectively. For $\Omega$ and $\omega$ they can be found in
(Iorio 2003). In order to have the precessions they must be
multiplied by $J_{2\odot}$. Note that the result for the mean
anomaly of Mercury accounts for the correction of order
$\mathcal{O}(e^2)$ which, instead, can be neglected for the other
planets. } \label{tab1}
$$
\begin{array}{p{0.5\linewidth}llllll}
            \hline
            \noalign{\smallskip}
Planet  & \dot\Omega_{\rm GE} & \dot\pi_{\rm GE} &
\dot\mathcal{M}_{\rm GE} & \dot\Omega_{.2}
& \dot\pi_{.2} & \dot\mathcal{M}_{.2} \\
\noalign{\smallskip}
            \hline
            \noalign{\smallskip}
Mercury & 0 & 42.981 & -127.949 & -126878.626 & 126404.437 & 123703.132 \\
Venus & 0 & 8.624 & -25.874 & -13068.273 & 13056.803 & 13056.504\\
Mars & 0 & 1.351 & -4.035 & -980.609 & 980.353 & 976.067\\
\noalign{\smallskip}
            \hline
         \end{array}
     $$
\end{table}
\begin{table}
\caption[]{Post-Newtonian gravito-magnetic precessions and
coefficients of Newtonian precessions of the node for Mercury,
Venus and Mars in $''$ cy$^{-1}$. The values $\dot\Omega_{\rm LT}$
are calculated with GTR. The coefficients $\dot\Omega_{.\ell}$,
are
$\partial(\dot\Omega_{J_{\ell\odot}})/\partial(J_{\ell\odot})$.
They can be found, up to degree $\ell=20$, in (Iorio 2003).
 In order to have the precessions they must be
multiplied by $J_{\ell\odot}$. } \label{tableti}
$$
\begin{array}{p{0.5\linewidth}lllll}
            \hline
            \noalign{\smallskip}
Planet  & \dot\Omega_{\rm LT} & \dot\Omega_{.2} & \dot\Omega_{.4} &  \dot\Omega_{\rm N-body}\\
\noalign{\smallskip}
            \hline
            \noalign{\smallskip}
Mercury & 0.001008 & -126878.626476 & 52.774935 & -446.30\\
Venus & 0.000144 & -13068.273031 & 1.349709 & -996.89\\
Mars & 0.000015 & -980.609460 & 0.023554 & -1020.19\\
\noalign{\smallskip}
            \hline
         \end{array}
     $$
\end{table}
In all the relativistic tests performed up to now by analyzing the
perihelia advances only of the inner planets of the Solar System
with the radar ranging technique (Shapiro et al. 1972; 1976;
Shapiro 1990) it has been impossible to disentangle the genuine
post-Newtonian gravito-electric contribution of \rfr{ge} from the
Newtonian precession of \rfr{obla}. Indeed, the observational
residuals of $\dot\pi$ for a single planet, built up by suitably
switching off the post-Newtonian $\mathcal{O}(c^{-2})$ terms and
the oblateness of Sun in the force models of the equations of
motion in the orbital processors softwares, account entirely for
the post-Newtonian and the solar oblateness\footnote{Also if the
solar oblateness is included in the force models, the related
uncertainty induces a corresponding systematic error in the
recovered post-Newtonian effect. Fortunately, it is small; by
assuming $\sigma_{J_{2\odot}}=0.4\times 10^{-7}$, it amounts to
0.01$\%$.} effects. This is a unsatisfactory situation, both if we
are interested in testing post-Newtonian gravity and if we want to
obtain a dynamical, model-independent measurement of $J_{2\odot}$.
Indeed, it is, of course, impossible to constraint both the
effects if only one perihelion rate is examined one at a time: in
recovering one of the two effects we are forced to consider the
other one as if it was known. Since the post-Newtonian
gravito-electric effect is three orders of magnitude larger than
that induced by solar oblateness, a determination of the latter by
assuming the validity of GTR would be affected by a non negligible
systematic error induced by the precision to which the
post-Newtonian pericentre advance is known from other (more or
less indirect and more or less biased by other aliasing effects)
tests (Lunar Laser Ranging, binary pulsars periastron
advance\footnote{Note that the binary pulsars periastron
measurement should not be considered as a test of relativistic
gravity because the masses of the binary system are not known
(Stairs et al. 1998). The Lunar Laser Ranging measurements do not
allow to single out uniquely the gravitoelctric pericentre advance
from the other post-Newtonian features of motion of the Earth-Moon
system (Nordvedt 2001). Recently, it has been proposed to measure
the relativistic gravito-electric perigee advance of the
terrestrial LAGEOS II satellite (Iorio et al. 2002b; Lucchesi
2003), but, up to now, the experiment has not yet been performed.
}).
%The problem is that,
%in the weak-field and slow-motion approximation of GTR valid
%throughout the Solar System, there are no tests of the
%post-Newtonian gravito-electric pericentre shifts other than those
%performed by measuring just the planetary perihelia advances.
The inverse situation is more favorable: indeed, if we are
interested in the post--Newtonian gravito-electric effect the
relative systematic error induced on its measurement by the
precession due to the solar oblateness amounts to $5\times
10^{-4}$ even by assuming for the latter effect a 100$\%$
uncertainty\footnote{Note that the impossibility of disentangle
the gravito-electric and Lense-Thirring effects would not
seriously affect the recovery of the gravitoelctric precession:
indeed, the bias induced by the gravito-magnetic effect on the
gravito-electric shift amounts to 0.004$\%$ only for Mercury.}.
\subsection{The present-day approach to test post-Newtonian gravity}
At this point it may be interesting to clarify what is the current
approach in testing post-Newtonian gravity from planetary data
analysis followed by, e.g., the Jet Propulsion Laboratory (JPL).
In the interplay between the real data and the equations of
motions, which include also the post-Newtonian accelerations
expressed in terms of the various PPN parameters, a set of
astrodynamical parameters, among which there are also $\gamma$ and
$\beta$, are simultaneously and straightforwardly fitted and
adjusted and a correlation matrix is also released. This means
that the post-Newtonian equations of motion are globally tested as
a whole in terms of, among other parameters, $\gamma$ and $\beta$;
no attention is paid to this or that particular feature of the
post-Newtonian accelerations. The point is that the standard PPN
formalism refers to the alternative theories of gravitation which
are metric, i.e. based on a symmetric spacetime metric. But it is
not proven that an alternative theory of gravitation must
necessarily be a metric one. Moreover, even in the framework of
the metric alternative theories, the PPN formalism based on 10
parameters is not sufficient to describe every conceivable metric
theory of gravitation at the post-Newtonian order; it only
describes those theories with a particularly simple post-Newtonian
limit. One would, in principle, need an infinite set of new
parameters to add to the standard ten parameter PPN formalism in
order to describe the post-Newtonian approximation of any a priori
conceivable metric theory of gravity (Ciufolini 1991; Ciufolini \&
Wheeler 1995).
\subsection{The possibilities opened by the future missions}\lb{newmiss}
Concerning the possibility of disentangle the effects of the solar
oblateness from those of the post-Newtonian gravito-electric
force, it is stated that the future space mission
BepiColombo\footnote{See on the WEB
\texttt{http://astro.estec.esa.nl/BepiColombo/}. Present ESA plans
are for a launch in 2010-2012.} of the European Space Agency (ESA)
will provide us, among other things, with a dynamical,
model-independent and relativity-independent measurement of
$J_{2\odot}$ by measuring with high precision the nodal motion of
Mercury (Milani et al. 2002; Pireaux \& Rozelot 2003) which is not
affected by the post-Newtonian gravito-electric force. The claimed
accuracy would amount to $\sigma_{J_{2\odot}}=2\times 10^{-9}$
(Milani et al. 2002). However, such evaluation refers to the
formal, statistical obtainable uncertainty only. Indeed, the
residuals of the Mercury's node would account, to a certain level
of accuracy, for the Lense-Thirring precession as well. By
considering such effect as totally unmodelled in the force models,
its impact on the measurement of $J_{2\odot}$ would induce a
$8\times 10^{-9}$ systematic error. The formal, statistical
accuracy for $\gamma$ and $\beta$ is evaluated to be of the order
of $2\times 10^{-6}$ (Milani et al. 2002). Also the ESA
astrometric mission GAIA\footnote{See on the WEB
\texttt{http://astro.estec.esa.nl/GAIA/}. Present ESA plans are
for a launch in mid-2010.} should measure, among other things, the
solar quadrupole mass moment by analyzing the longitudes of the
ascending nodes of many minor bodies of the Solar System. The
obtainable accuracy for $\gamma$ is of the order of
$10^{-5}$-10$^{-7}$ (Vecchiato et al. 2003). The ASTROD mission
should be able to measure $J_{2\odot}$ with a claimed accuracy of
the order of $10^{-8}$ or, perhaps, $10^{-9}-10^{-10}$ (Ni et al.
2004). The claimed obtainable accuracy for the PPN parameters is
$4.6\times 10^{-7}$ for $\gamma$ and $4\times 10^{-7}$ for
$\beta$. Further improvements may push these limits down to
$10^{-8}-10^{-9}$. The recently proposed LATOR mission should be
able to measure, among other things, $\gamma$ to a 10$^{-8}$
accuracy level and $J_{2\odot}$ to a $10^{-8}$ level (Turyshev et
al. 2004). However, LATOR, whose goal is an extremely accurate
measurement of the light deflection at the Sun's limb, should not
allow to disentangle the various post-Newtonian contributions
(first and second order gravito-electric contributions, first
order gravito-magnetic term, solar oblateness effect) to it. The
solar orbit relativity test SORT (Melliti et al. 2002), which
would combine a time-delay experiment with a light deflection
test, should allow to reach a $10^{-6}$ accuracy in measuring the
PPN parameters
\section{The observational accuracy in planetary radar ranging}
Concerning the present-day accuracy of the planetary radar
ranging, the radar itself is accurate well below the 100 m level
(Standish 2002). The problem, however, comes from the fact that
the surfaces of the planets have large topographical variations.
They are modeled in different ways. For Mercury, spherical
harmonics and some closure analysis (comparing values when two
different measurements reflect off from the same spot on the
surface) have been done (Anderson et al. 1996) in DE405. For
Venus, a topographical model, which comes from (Pettengill et al.
1980), has been used. For Mars, closure points can be used.
Closure points are pairs of days during which the observed points
on the surface of Mars are nearly identical with respect to their
longitudes and latitudes on Mars. Since the same topographical
features are observed during each of the two days, the uncertainty
introduced by the topography may be eliminated by subtracting the
residuals of one day from the corresponding ones of the other day.
The remaining difference is then due to only the ephemeris drift
between the two days. The closure points for Mars have a priori
uncertainties of about 100 m or less when the points are within
0.2 degrees of each other on the martian surface. Of course, for
Mars, there is also the spacecraft ranging - far more accurate
than the radar: Viking Landers (1976-82), 10 m; Mars Global
Surveyor and Odyssey (1999-2003), 2-3 m. However, correction for
Mars topography is possible not only by using closure points (in
this method no all observations may be used), but with help of
modern hypsometric maps and by the representations of the global
topography with an expansion of spherical functions (Pitjeva
2001b).
\begin{table}
\caption[]{Present-day accuracy in determining the node,
perihelion and mean anomaly secular rates of Mercury, Venus and
Mars according to DE405 (Standish 1998) and EPM2000 (Pitjeva
2001a; 2001b) ephemerides. The figures, in $''$ cy$^{-1}$,
represent the formal, statistical errors. Realistic errors should
be 10 times larger, at least. While the results by Standish come
from the mathematical propagation  of the nodes, the perihelia and
the mean anomalies (Standish, E. M. private communication)
evolution with and without post--Newtonian terms (with
$\gamma=\beta=1$) and their average over a time span of two
centuries, the results by Pitjeva are based on real data. The
figure for Mercury has been obtained in 2001, while the other ones
have been determined subsequently (Pitjeva, personal communication
2004). } \label{rates}
$$
         \begin{array}{p{0.5\linewidth}llll}
            \hline
            \noalign{\smallskip}
Planet  & \sigma_{\dot\Omega_{\rm calc}} (DE405) &
\sigma_{\dot\pi_{\rm calc}} (DE405) & \sigma_{\dot\mathcal{M}_{\rm
calc}}
(DE405) & \sigma_{\dot\pi_{\rm obs}} (EPM2000) \\
\noalign{\smallskip}
            \hline
            \noalign{\smallskip}
Mercury & 0.000182 & 0.0023 & 0.0027 & 0.0086\\
Venus & 0.000006 & 0.0414 & 0.0414 & 0.1037\\
Mars & 0.000001 & 0.0014 & 0.0014 & 0.0001\\
\noalign{\smallskip}
            \hline
         \end{array}
     $$
\end{table}
\section{The proposed approach for disentangle the solar oblateness and the post-Newtonian effects}
In regard to the possibility of measuring explicitly some
post-Newtonian effects from the analysis of the orbital motion of
proof masses in the gravitational field of a real (rotating)
astronomical mass like the Earth, it must be pointed out that the
main problems come from the aliasing effects induced by a host of
classical orbital perturbations, of gravitational and
non-gravitational origin, which unavoidably affect the motion of
the probes along with GTR. In particular, the even zonal harmonics
$J_{\ell}$ of the multipolar expansion of the gravitational
potential of the central mass induce secular classical precessions
which, in many cases, tend to alias the gravito-electromagnetic
ones of interest. Moreover, also the non-gravitational
perturbations, to which the perigees of the LAGEOS--like
satellites are particularly sensitive,  are another important
source of bias. The approach proposed by Ciufolini (1996) and
Iorio (Iorio 2002; Iorio \& Morea 2004) in the performed or
proposed tests with LAGEOS and LAGEOS II consists of suitably
designing linear combinations $\sum u_i\delta\dot\Omega^i_{\rm
obs}+\sum v_j\delta\dot\omega^j_{\rm obs}$ of orbital residuals
which are able to reduce the impact both of the even zonal
harmonics of the gravitational field of the central mass and of
the non--gravitational perturbations. In general, the coefficients
$u_i$ and $v_j$ which weigh the various orbital elements in the
combinations are a compromise between these two distinct needs.
\subsection{The gravito-electric precessions and the solar oblateness}
The main idea is to extend this approach to the Sun-planets
system. It must considered that, in this case, the
non-gravitational perturbations do not play any
role\footnote{Indeed, they are proportional to the area-to-mass
ratio $S/M$ of the orbiting probe. The area-to-mass ratio falls
off as $1/r$ where $r$ is the radius of the probe-a planet, in
this case-assumed spherical.}, while the $N-$body classical
perturbations of gravitational origin are important. Let us start
with the gravito-electric precessions writing down the following
equations\footnote{Note that in the right-hand-sides of
\rfrs{uno}{due} also the mismodelled parts of the classical
$N-$body precessions should have been included. Since they are
small with respect to the gravito-electric effects of interest, as
we will show later, we can neglect them in the calculations for
setting up our combiantions. This, of course, does not mean that
the residuals do not account also for them.}

\begin{equation}
\left\{
\begin{array}{lll}\lb{geidue}
\delta\dot\Omega^{\rm Merc}_{\rm obs}&=&\dot\Omega^{\rm
Merc}_{.2}J_{2\odot}+\dot\Omega^{\rm Merc}_{\rm
N-body}+\dot\Omega^{\rm Merc}_{\rm LT}\mu_{\rm
LT},\\\\
\delta\dot\Omega^{\rm Venus}_{\rm obs}&=&\dot\Omega^{\rm
Venus}_{.2}J_{2\odot}+\dot\Omega^{\rm Venus}_{\rm
N-body}+\dot\Omega^{\rm Venus}_{\rm LT}\mu_{\rm
LT},\\\\
\delta\dot\Omega^{\rm Mars}_{\rm obs}&=&\dot\Omega^{\rm
Mars}_{.2}J_{2\odot}+\dot\Omega^{\rm Mars}_{\rm
N-body}+\dot\Omega^{\rm Mars}_{\rm LT}\mu_{\rm
LT},\\\\
\end{array}
\right.
\end{equation}

\begin{equation}
\left\{
\begin{array}{lll}\lb{uno}
\delta\dot\mathcal{M}^{\rm Mars}_{\rm obs}&=&\dot\mathcal{M}^{\rm
Mars}_{.2}J_{2\odot}+\dot\mathcal{M}^{\rm Mars}_{\rm GE}\mu_{\rm
GE},\\\\
\delta\dot\mathcal{M}^{\rm Venus}_{\rm obs}&=&\dot\mathcal{M}^{\rm
Venus}_{.2}J_{2\odot}+\dot\mathcal{M}^{\rm Venus}_{\rm GE}\mu_{\rm
GE},\\
\end{array}
\right.
\end{equation}

\begin{equation}
\left\{
\begin{array}{lll}\lb{due}
\delta\dot\pi^{\rm Mars}_{\rm obs}&=&\dot\pi^{\rm
Mars}_{.2}J_{2\odot}+\dot\pi^{\rm Mars}_{\rm GE}\nu_{\rm
GE},\\\\
\delta\dot\pi^{\rm Merc}_{\rm obs}&=&\dot\pi^{\rm
Merc}_{.2}J_{2\odot}+\dot\pi^{\rm Merc}_{\rm GE}\nu_{\rm
GE},\\
\end{array}
\right.
\end{equation}
where $\delta\dot\Omega^{\rm Planet}_{\rm obs}$,
$\delta\dot\pi^{\rm Planet}_{\rm obs}$ and
$\delta\dot\mathcal{M}^{\rm Planet}_{\rm obs}$ are the
observational residuals of the rates of the nodes, the longitudes
of the perihelia and the mean anomalies of Mercury, Venus and
Mars. It is intended that all kind of data (optical and radio)
would be used. The residuals\footnote{Note that here we use
expressions like `observational residuals of a Keplerian orbital
element', strictly speaking, in an improper sense. The true
observables, for which it is possible to build up residuals, are
ranges, range rates and angles. The Keplerian orbital elements are
not directly observable: they can only be calculated from an orbit
which has been either numerically propagated or fitted to real
observational data (the observed orbit) starting from given
initial conditions. Here with `time series of the residuals of a
Keplerian element' we mean the difference between the time series
of that Keplerian element related to an observed orbital arc and
the time series of that Keplerian element numerically propagated
on the same orbital arc (equal initial conditions) with the given
classical or post-Newtonian parameter of interest switched off in
the force models.} should be built up by purposely switching off
the solar quadrupole moment and the post-Newtonian
gravito-electric accelerations in the force models (or leaving in
them some default values to be subsequently adjusted according to
the present strategy) of the orbital processors. Then, the so
obtained observational residuals would entirely (or partly, if
some default values are left in the force models) adsorb just the
investigated secular effects and other post--Newtonian
short--periodic features, i.e. not averaged over one orbital
revolution of the planet under consideration. In respect to the
latter point, it should be noted that the residuals of the mean
anomalies would account, e.g., also for the indirect effects on
the mean motions $n$ through the perturbations in the semimajor
axes $a$ \eqi\Delta n=-\rp{3}{2}\sqrt{\rp{GM}{a^5}}\Delta
a\lb{deltan}.\eqf  There are no secular perturbations induced on
$a$ by the other planets. If the classical short--periodic effects
on $a$ would be of relatively no importance because they would be
included in the force models at the best of their accuracy, this
is not the case for the post-Newtonian ones. The gravito-electric
field induces no secular variations on $a$, as the classical
planetary perturbations. The short--term shift on $a$ can be
calculated from \rfr{age} and the Gauss equation for the perturbed
rate of semimajor axis \eqi
\frac{da}{dt}=\rp{2}{n\sqrt{1-e^2}}\left[A_R e\sin f+A_T(1+e\cos f
)\right], \eqf where $A_T$ is the along-track component of the
perturbing acceleration. It amounts to \eqi \Delta a_{\rm
GE}=\rp{GMe}{c^2 (1-e^2)^2}[14(\cos f_0-\cos f)+10e(\cos^2
f-\cos^2 f_0 )]+\mathcal{O}(e^3).\eqf For Venus their amplitudes
are of the order of 100 m; for Mars they amount to 2 km;
\rfr{deltan} would yield periodic variations whose amplitudes
would be of the order of 0.2-1 $''$ cy$^{-1}$ for Venus and Mars,
respectively. However, such harmonic signatures could be fitted
and removed from the data over sufficiently long time spans.
%It turns out that the indirect effects of $\Delta a_{\rm GE
%}$ on the planetary perturbations on $\dot\mathcal{M}$ are
%negligible.
%; indeed, an explicit calculation of the
%cross--coupling between the shift of the semimajor axis of the
%planet $m_i$ and the secular planetary perturbations induced on
%$m_i$ by the planet $m_j$ yields, for Mercury and Jupiter, a rate
%of $10^{-5}$ $''$ cy$^{-1}$.
Indeed, since we are interested in the gravito-electric secular
trends on $\pi$ and $\mathcal{M}$ it should be possible, in
principle, to construct the residuals by using orbital arcs longer
than the sidereal revolution periods of the planets to be used.
Then, all the high--frequency perturbations would not affect their
time series which should, instead, be characterized by the secular
parts of those Newtonian and post--Newtonian features present in
the real data but (partly) absent in the force models of the
equations of motion in the orbital processors. However, in regard
to the possibility of constructing accurate time series of
observational residuals many years long the following observations
must be kept in mind (Standish 2002). The planetary motions are
perturbed by the presence of many asteroids whose masses are quite
poorly known. Furthermore, it is not possible to solve for the
asteroid masses, other than for the biggest few, because there are
too many of them for the data to support such an effort. As a
result, the ephemerides of the inner planets, especially Mars,
will deteriorate over time; the ephemerides have uncertainties at
the 1-2 km level over the span of the observations and growing at
the rate of a few km/decade outside that span. On the other hand,
it must also be noted that the sidereal orbital periods of Mercury
and Mars amount to 87.969 and 686.980 days, respectively.

We can consider \rfrs{geidue}{due} as three systems of algebraic
linear equations in the three unknowns\footnote{This approach is
analogous to that employed in the LAGEOS-LAGEOS II Lense-Thirring
experiment in the gravitational field of Earth (Ciufolini 1996;
Ciufolini et al. 1998)} $J_{2\odot}, \mu_{\rm GE}$ and $\nu_{\rm
GE}$. Their solutions can be written as
\begin{equation}
\left\{
\begin{array}{lll}\lb{tre}
\delta\dot\Omega^{\rm Merc}_{\rm obs}+c_1\delta\dot\Omega^{\rm
Venus}_{\rm obs}+c_2\delta\dot\Omega^{\rm Mars}_{\rm obs}&=&
J_{2\odot}\left(\dot\Omega^{\rm
Merc}_{.2}+c_1\dot\Omega^{\rm Venus}_{.2}+c_2\dot\Omega_{.2}^{\rm Mars}\right),\\\\
\delta\dot\mathcal{M}^{\rm Mars}_{\rm
obs}+c^{'}_1\delta\dot\mathcal{M}^{\rm Venus}_{\rm obs}&=&\mu_{\rm
GE}\left(\dot\mathcal{M}^{\rm
Mars}_{\rm GE}+c^{'}_1\dot\mathcal{M}^{\rm Venus}_{\rm GE}\right),\\\\
\delta\dot\pi^{\rm Mars}_{\rm obs}+c^{''}_1\delta\dot\pi^{\rm
Merc}_{\rm obs}&=& \nu_{\rm GE}\left(\dot\pi^{\rm
Mars}_{\rm GE}+c^{''}_1\dot\pi^{\rm Merc}_{\rm GE}\right),\\
\end{array}
\right.
\end{equation}
where
\begin{equation}
\left\{
\begin{array}{lll}\lb{quattro}
c_1&=&\frac{\dot\Omega^{\rm Mars}_{\rm LT}\dot\Omega^{\rm
Merc}_{\rm N-body } -\dot\Omega^{\rm Merc}_{\rm LT}\dot\Omega^{\rm
Mars}_{\rm N-body }}{\dot\Omega^{\rm Venus}_{\rm
LT}\dot\Omega^{\rm Mars}_{\rm N-body }-\dot\Omega^{\rm Mars}_{\rm
LT}\dot\Omega^{\rm Venus}_{\rm N-body}}=-7.73247,\

c_2=\frac{\dot\Omega^{\rm Merc}_{\rm LT}\dot\Omega^{\rm
Venus}_{\rm N-body} -\dot\Omega^{\rm Venus}_{\rm
LT}\dot\Omega^{\rm Merc}_{\rm N-body }}{\dot\Omega^{\rm
Venus}_{\rm LT}\dot\Omega^{\rm Mars}_{\rm N-body }-\dot\Omega^{\rm
Mars}_{\rm LT}\dot\Omega^{\rm Venus}_{\rm
N-body}}=7.11840\\\\
c^{'}_1&=&-{\dot\mathcal{M}^{\rm Mars}_{.2}}/{\dot\mathcal{M}^{\rm Venus}_{.2}}=-0.07475,\\\\
c^{''}_1&=&-{\dot\pi^{\rm Mars}_{.2}}/{\dot\pi^{\rm Merc}_{.2}}=-0.00775,\\
\end{array}
\right.
\end{equation}
and
\begin{equation}
\left\{
\begin{array}{lll}\lb{cinque}
\dot\Omega^{\rm Merc}_{.2}+c_1\dot\Omega^{\rm
Venus}_{.2}+c_2\dot\Omega^{\rm Mars}_{.2}&=&-32808.8816\ ''\ {\rm cy}^{-1},\\\\
\dot\mathcal{M}^{\rm Mars}_{\rm GE}+c^{'}_1\dot\mathcal{M}^{\rm
Venus}_{\rm GE}&=& -2.1007\ ''\ {\rm
cy}^{-1},\\\\
\dot\pi^{\rm Mars}_{\rm GE}+c^{''}_1\dot\pi^{\rm Merc}_{\rm
GE}&=&1.0176\ ''\ {\rm cy}^{-1}\\
\end{array}
\right.
\end{equation}
The first equation of \rfr{tre} comes from \rfr{geidue} solved for
$J_{2\odot}$; it allows to obtain $J_{2\odot}$ independently of
the post-Newtonian Lense-Thirring and classical $N-$ body secular
precessions which would represent the major sources of systematic
errors. The second equation of \rfr{tre} comes from \rfr{uno}
solved for $\mu_{\rm GE}$; it cancels out the secular precessions
due to $J_{2\odot}$. The third equation in \rfr{tre} comes from
\rfr{due} solved for $\nu_{\rm GE }$; it cancels out the secular
precessions due to $J_{2\odot}$. In regard to the impact of the
residual mismodelled classical $N-$ body precessions on the second
and the third combinations of \rfr{tre}, they should not induce a
systematic error larger than the observational one (see below)
because the expected values of $\mu_{\rm GE}$ and $\nu_{\rm GE}$
are of the order of unity\footnote{Let us quantitatively discuss
this point. By using the results for the observed centennial rates
$\dot\Omega$ and $\dot\omega$ released at
\texttt{http://ssd.jpl.nasa.gov/elem$\_$planets.html} in, say, the
left-hand-side of the third combination of \rfr{tre} it is
possible to obtain for it a nominal $N-$body shift of 539.6036
$''$ cy$^{-1}$. The uncertainty in the $N-$body precessions lies
mainly in the $Gm$ of the perturbing planets, among which Jupiter
plays the major role. Now, the relative uncertainty in Jupiter's
$Gm$ is of the order of 10$^{-8}$ (Jacobson 2003); then, a
reasonable estimate of the order of magnitude of the mismodelled
part of the $N-$body shift should be $1\times 10^{-5}$ $''$
cy$^{-1}$. This figure must be divided by 1.0176 $''$ cy$^{-1}$
yielding a relative error in $\nu_{\rm GE}$ of the order of
$10^{-5}$. Cfr. with the third equations of \rfr{sette}.},
contrary to $J_{2\odot}$ which should be of the order of
$10^{-7}$.  The adimensional parameters $J_{2\odot},\ \mu_{\rm
GE}$ and $\nu_{\rm GE}$ are estimated by fitting the time series
of the left-hand-sides of \rfr{tre} with straight lines, measuring
their slopes, in $''$ cy$^{-1}$, and, then, by dividing them by
the the quantities of \rfr{cinque} which have the dimensions of
$''$ cy$^{-1}$. Note that the solar quadrupole mass moment would
not be affected by the indirect effects on $n$ because only the
nodes would be used in its determination. Finally, from the so
obtained values of $\mu_{\rm GE}$ and $\nu_{\rm GE}$, which are 1
in GTR and 0 in Newtonian mechanics, it is possible to measure
$\gamma$ and $\beta$ independently of the solar oblateness and
also of each other as
\begin{equation}
\left\{
\begin{array}{lll}\lb{sei}
\gamma&=&\frac{9}{10}(\mu_{\rm GE}+\nu_{\rm GE})-\frac{4}{5},\\\\
\beta&=&\rp{9}{5}\mu_{\rm GE}-\rp{6}{5}\nu_{\rm GE}+\rp{2}{5}.\\
\end{array}
\right.
\end{equation}
According to the results of Table \ref{rates} it is possible to
yield an estimate of the (formal) uncertainty in $J_{2\odot},\
\mu_{\rm GE}$ and $\nu_{\rm GE}$ as
\begin{equation}
\left\{
\begin{array}{lll}\lb{sette}
\sigma_{J_{2\odot}}&\sim& 4.5\times 10^{-9},\\\\
\sigma_{\mu_{\rm GE}}&\sim& 6\times 10^{-5},\\\\
\sigma_{\nu_{\rm GE}}&\sim& 3\times 10^{-5}.\\
\end{array}
\right.
\end{equation}
The evaluation for $\sigma_{\mu_{\rm GE}}$ accounts also for the
fact that \rfr{deltan} and the values of $\sigma_a$ of Table VI of
(Pitjeva 2001a) yield a (formal) uncertainty of $3\times 10^{-5}$.
The uncertainties in $\gamma$ and $\beta$ would amount to
\begin{equation}
\left\{
\begin{array}{lll}\lb{otto}
\sigma_{\gamma}&\sim& 6\times 10^{-5},\\\\
\sigma_{\beta}&\sim& 1\times 10^{-4}.\\
\end{array}
\right.
\end{equation}
However, discretion is advised on evaluating the reliability of
these results because they refer to the formal, standard
statistical errors; realistic errors may be also one order of
magnitude larger.

For the most recent determinations of $\gamma$ and $\beta$ we have
that, according to the frequency shift of radio photons to and
from the Cassini spacecraft, $\sigma_{\gamma}=2\times 10^{-5}$
(Bertotti et al. 2003). This result, combined with that for the
Strong Equivalence Principle violating parameter
$\sigma_{\eta}=4.5\times 10^{-4}$ from the most recent analysis of
LLR data, yields $\sigma_{\beta}=1\times 10^{-4}$ (Turyshev et al.
2003).
\subsection{The gravito-magnetic precessons}
Let us, now, turn our attention to the Lense-Thirring effect and
try to set up some combinations with the nodes of the inner
planets which cancel out the impact of the Sun's oblateness and of
the $N-$ body precessions.
\subsubsection{A $J_{2\odot}-J_{4\odot}$ free
combination} Let us write down the expressions of the residuals of
the nodes of Mercury, Venus and Mars explicitly in terms of the
mismodelled secular precessions induced by the quadrupolar and
octupolar (Rozelot et al. 2004) mass moments of the Sun and of the
Lense-Thirring secular precessions, assumed as a totally
unmodelled feature. It is accounted for by a scaling parameter
$\mu_{\rm LT}$ which is zero in Newtonian mechanics and 1 in GTR
%It is intended that the
%time series of the residuals are averaged over one orbital
%revolution of the considered planets
\begin{equation}\left\{\begin{array}{lll}
\delta\dot\Omega^{\rm Mercury}_{\rm obs}=\dot\Omega^{\rm
Mercury}_{.2} \delta J_{2\odot}+\dot\Omega_{.4}^{\rm Mercury}
\delta J_{4\odot}+\dot\Omega^{\rm Mercury}_{\rm LT}\mu_{\rm LT}+\Delta^{\rm Mercury},\\\\
\delta\dot\Omega^{\rm Venus}_{\rm obs}=\dot\Omega^{\rm Venus}_{.2}
\delta J_{2\odot}+\dot\Omega^{\rm Venus}_{.4}
\delta J_{4\odot}+\dot\Omega^{\rm Venus}_{\rm LT}\mu_{\rm LT}+\Delta^{\rm Venus},\\\\
\delta\dot\Omega^{\rm Mars}_{\rm obs}=\dot\Omega^{\rm Mars}_{.2}
\delta J_{2\odot}+\dot\Omega_{.4}^{\rm Mars} \delta
J_{4\odot}+\dot\Omega^{\rm Mars}_{\rm LT}\mu_{\rm LT}+\Delta^{\rm
Mars}.\label{syst}
\end{array}\right.\end{equation}
The coefficients $\dot\Omega_{.{\ell}}$ for the first two even
zonal harmonics are
\begin{equation}\left\{\begin{array}{lll}\label{precclass}
\dot\Omega_{.2}=-\frac{3}{2}n\left(\frac{R_{\odot}}{a}\right)^2\frac{\cos
i }{(1-e^2)^2},\\\\
\dot\Omega_{.4}=\dot\Omega_{.2}\left[\frac{5}{8}\left(\frac{R}{a}\right)^2\frac{\left(1+\frac{3}{2}e^2\right)}{(1-e^2)^2}
\left(7\sin^2 i-4\right)\right]
\end{array}\right.\end{equation}
$R_{\odot}$ is the Sun's equatorial radius. The quantities
$\Delta$ in \rfr{syst} refer to the other unmodelled or
mismodelled effect which affect the temporal evolution of the
nodes of the considered planets. In the present case they would be
the $N-$body gravitational interactions$-$of Newtonian
origin$-$with the other planets and asteroids of the Solar System.
From the nominal values of $\dot\Omega_{\rm N-body}$ quoted in
Table \ref{tableti} it can be noted that their secular components
are very important perturbing effects which should be known with
very high accuracy in order to prevent a fatal aliasing effect on
the proposed measurement.

If we solve \rfr{syst} for the Lense-Thirring parameter $\mu_{\rm
LT }$ it is possible to obtain
\begin{equation}\left\{\begin{array}{lll}\delta\dot\Omega_{\rm obs}^{\rm
Mercury}  +  k_1\delta\dot\Omega_{\rm obs}^{\rm
Venus}+k_2\delta\dot\Omega_{\rm obs}^{\rm Mars}=X_{\rm LT}\mu_{\rm
LT}+({\rm other\ secular\ mismodelled\ effects\ \Delta}),\\\\
k_1=\frac{\dot\Omega^{\rm Mars}_{.2}\dot\Omega^{\rm Mercury}_{.4}
-\dot\Omega^{\rm Mercury}_{.2}\dot\Omega^{\rm
Mars}_{.4}}{\dot\Omega^{\rm Venus}_{.2}\dot\Omega^{\rm
Mars}_{.4}-\dot\Omega^{\rm Mars}_{.2}\dot\Omega^{\rm Venus}_{.4}}=
-48.008308,\\\\
 k_2=\frac{\dot\Omega^{\rm Mercury}_{.2}\dot\Omega^{\rm Venus}_{.4}
-\dot\Omega^{\rm Venus}_{.2}\dot\Omega^{\rm
Mercury}_{.4}}{\dot\Omega^{\rm Venus}_{.2}\dot\Omega^{\rm
Mars}_{.4}-\dot\Omega^{\rm Mars}_{.2}\dot\Omega^{\rm
Venus}_{.4}}=510.404066,\\\\X_{\rm LT}=\dot\Omega_{\rm LT }^{\rm
Mercury}+k_1\dot\Omega_{\rm LT}^{\rm Venus}+k_2\dot\Omega^{\rm
Mars }_{\rm LT }=0.002069\ '' \ {\rm cy}^{-1},
\label{ltformula}\end{array}\right.\end{equation} where the
numerical values of the coefficients $k_1$ and $k_2$ and the slope
$X_{\rm LT}$ of the gravito-magnetic trend come from the values of
Table \ref{tableti}. The meaning of \rfr{ltformula} is the
following. Let us construct the time series of the residuals of
the nodes of Mercury, Venus and Mars by using real observational
data and the full dynamical models in which GTR is purposely set
equal to zero, e.g. by using a very large value of $c$. We expect
that, over a multidecadal observational time span, the so combined
residuals will fully show the GTR signature and partly the
mismodelled $N-$body effect\footnote{Possible aliasing
time-dependent $N-$body residual effects with the periodicities of
the outer planets, mainly Jupiter, should average out over a
sufficiently long multidecadal time span. However, it would be
possible to fit and remove them from the time series.}in terms of
a linear trend. The measured slope, divided by $X_{\rm LT}$,
yields $\mu_{\rm LT}$ which should be equal to one if GTR was
correct and if the bias from the residual $N-$body effect was
sufficiently small. If we use the results of Table \ref{rates} in
\rfr{ltformula} in order to get an idea of what could be the
observational accuracy of such a measurement we get a 19$\%$
error. It is important to note that such an estimate does not
include the systematic error induced by mismodelled secular effect
due to the classical $N-$body interactions.
\subsubsection{A
$J_{2\odot}-(N-$body$)$ free combination} \Rfr{ltformula} is
designed in order to cancel out the contributions of the first two
even zonal harmonics of the multipolar expansion of the Sun's
gravitational potential. However, it is affected by the $N-$body
classical secular precessions which, as it turns out from the
figures in Table \ref{tableti}, are quite large; their mismodelled
part, which would be accounted for by the orbital residuals, could
be turn out to be very insidious with respect to our proposed
measurement of the Lense-Thirring effect.

In view of these considerations, it would be better to set up
another observable which is insensitive just to such huge
classical secular effects. Moreover, from the results of Table
\ref{tableti} and from the evaluations of (Rozelot et al. 2004)
according to which the possible magnitude of $J_{4\odot}$ would
span the range $10^{-7}-10^{-9} $, the secular precession induced
by the octupolar mass moment of the Sun is negligible with respect
the Lense-Thirring rates. These facts lead us to design a
three-node combination which cancels out the effects of
$J_{2\odot}$ and of the classical $N-$body precessions and is
affected by $J_{4\odot}$. It can be obtained by substituting
$\dot\Omega_{.4}$ with $\dot\Omega_{\rm class}$ in
\rfr{ltformula}. It is
\begin{equation}\left\{\begin{array}{lll}\delta\dot\Omega_{\rm obs}^{\rm
Mercury}  +  k^{'}_1\delta\dot\Omega_{\rm obs}^{\rm
Venus}+k^{'}_2\delta\dot\Omega_{\rm obs}^{\rm Mars}\sim Y_{\rm
LT}\mu_{\rm
LT},\\\\
k^{'}_1=\frac{\dot\Omega^{\rm Mars}_{.2}\dot\Omega^{\rm
Mercury}_{\rm class } -\dot\Omega^{\rm
Mercury}_{.2}\dot\Omega^{\rm Mars}_{\rm class }}{\dot\Omega^{\rm
Venus}_{.2}\dot\Omega^{\rm Mars}_{\rm class }-\dot\Omega^{\rm
Mars}_{.2}\dot\Omega^{\rm Venus}_{\rm class}}=-10.441702
,\\\\
 k^{'}_2=\frac{\dot\Omega^{\rm Mercury}_{.2}\dot\Omega^{\rm Venus}_{\rm class}
-\dot\Omega^{\rm Venus}_{.2}\dot\Omega^{\rm Mercury}_{\rm class
}}{\dot\Omega^{\rm Venus}_{.2}\dot\Omega^{\rm Mars}_{\rm class
}-\dot\Omega^{\rm Mars}_{.2}\dot\Omega^{\rm Venus}_{\rm
class}}=9.765758,\\\\Y_{\rm LT}=\dot\Omega_{\rm LT }^{\rm
Mercury}+k^{'}_1\dot\Omega_{\rm LT}^{\rm
Venus}+k^{'}_2\dot\Omega^{\rm Mars }_{\rm LT }=-0.000351\ '' \
{\rm cy}^{-1}, \label{ltformula2}\end{array}\right.\end{equation}
The systematic error affecting \rfr{ltformula2} should be totally
negligible because it would be due only to the higher degree
multipole mass moments of the Sun. The observational error,
according to the results of Table \ref{rates}, would amount to
36$\%$.
\section{Conclusions}
In this chapter we have proposed a relatively cheap  approach in
testing post-Newtonian gravity in the Solar System scenario with
particular emphasis on some secular gravito-electromagnetic
effects on the orbital motion of the planets. It is based on a
reanalysis and a suitable combination of the ranging data to
planets. As a by-product, also the important solar quadrupole mass
moment $J_{2\odot}$ could be accurately measured.

First, we have explicitly worked out the post--Newtonian
gravito-electric secular rate of the mean anomaly of a test
particle freely orbiting a spherically symmetric central object.
Moreover, we have outlined a possible strategy for determining
simultaneously and independently of each other, the solar
quadrupole mass moment $J_{2\odot}$ and two parameters $\nu_{\rm
GE}$ and $\mu_{\rm GE}$ which account for the post--Newtonian
gravito-electric secular shifts of the perihelion and the mean
anomaly, respectively, of planets. They are 0 in Newtonian
mechanics and 1 in the General Theory of Relativity. They could be
expressed in terms of the standard PPN $\gamma$ and $\beta$
parameters which could, then, be determined independently of each
other as well. The usual approach employed in the ephemerides data
reductions tests post--Newtonian gravity theories as a whole
straightforwardly in terms of the PPN parameters involving the
simultaneous fit of many more or less correlated astrodynamical
parameters among which there are also $\gamma$, $\beta$ and
$J_{2\odot}$. In this case, instead, we propose to analyze the
time series of three suitably linearly combined residuals of
$\dot\Omega$, $\dot\pi$ and $\dot\mathcal{M}$ of Mercury, Venus
and Mars built up in order to single out just certain selected
Newtonian and post--Newtonian orbital effects. This can be done by
setting purposely equal to zero (or to some default values to be
subsequently adjusted with the proposed strategy) the orbital
effects of interest in the force models of the equations of
motion. By suitably choosing the length of the orbital arcs it
would be possible to account for the secular terms only. The
coefficients of the combinations would make each of them sensitive
just to one orbital effect, independently of the other ones. By
fitting the experimental residual signals with straight lines,
measuring their slopes in $''$ cy$^{-1}$ and suitably normalizing
them would yield the values of $J_{2\odot},\ \mu_{\rm GE}$ and
$\nu_{\rm GE}$. The obtainable formal accuracy would be of the
order of $10^{-4}-10^{-5}$ for $\gamma$ and $\beta$ and, more
interestingly, 10$^{-9}$ for $J_{2\odot}$.

Second, we have extended this strategy to the gravito-magnetic
Lense-Thirring effect induced by the Sun's angular momentum. We
have designed two linear combinations which would allow to measure
such tiny effect, by construction, independently of the aliasing
impact of the Sun's oblateness and of the $N-$body precessions.
The obtainable accuracy would be of the order of 20-30$\%$.

However, it must be pointed out that many of the proposed tests
involve the analysis of the orbital motion of Mercury. The quality
and the accuracy of its ephemerides should increase in future
thanks to the already launched Messenger and planned BepiColombo
spacecrafts, so that it is reasonably presumable that the
precision of the proposed tests will increase as well.
%-----------------------------------------
\begin{acknowledgements}
L. Iorio is grateful to L. Guerriero for his support while at
Bari. Special thanks to E. M. Standish and E. Pitjeva for their
help and useful discussions and clarifications, and to W.-T. Ni
for the updated reference about ASTROD.
\end{acknowledgements}
%-----------------------------------------

%-----------------------------------------------------------

\end{document}